\documentclass{PoS}

\title{Cross-disciplinary research in astronomy}

\ShortTitle{Cross-disciplinary astronomy}

\author{\speaker{Eric D. Feigelson}\\
        Department of Astronomy \& Astrophysics, Pennsylvania State University,
        525 Davey Laboratory, University Park PA 16802 U.S.A. \\
        E-mail: \email{edf@astro.psu.edu}}

\abstract{In the distant past, astronomy was often intertwined with religion into a unified cosmos. As science became a distinct cultural enterprise, astronomy has witnessed a variety of rich interactions with other fields.  Mathematical statistics was stimulated in the 19th century by astronomical problems, and today astrostatistics is a small but growing cross-disciplinary field advancing methodology to address challenges in astronomical data analysis.  Throughout the 20th century, astronomy became closely allied with physics such that astronomy and astrophysics are now profoundly intertwined.  Physical chemistry played a major role in the identification of molecules in the Milky Way Galaxy, and astrochemistry is now an active subfield giving insights into cosmic molecular processes.  The importance of cross-disciplinary interactions with engineering (for instrumentation), Earth sciences (for planetary studies), computer science (for astroinformatics) and life sciences (for astrobiology) is also growing.  Cross-disciplinary research has been essential both for crucial discoveries in astronomy and for improving the quality of astronomical research.  It should be fostered with increased flexibility in the training of young astronomers and with sufficient funding to nurture these fields.  
}

\FullConference{Accelerating the Rate of Astronomical Discovery, sps5\\
		August 11-14, 2009\\
		Rio de Janeiro, Brazil}

\begin{document}

\section{The close integration of astronomy and culture in early societies}

In many ancient and long-lived civilizations, astronomy was a cross-disciplinary enterprise in the broadest sense.  The interpretation of celestial bodies and motions was an integral part of the worldview with little differentiation between (what we today call) science and religion. In Babylonia, astronomical bodies were simultaneously associated with the measures of time, with deities, and with human affairs.  Remnants of these relationships still remain in the English language in the days of the week.  Monday, Friday and Saturday were associated with the Moon, Venus and Saturn respectively, which in turn were associated with the deities Sin, Ishtar and Ninurta.  The words "Monday" and "Saturday" retain the link directly, while "Friday" is named after Frigg, the Germanic goddess of Venus.  For the Babylonians, the astronomical deities were directly involved in both global events like famines or floods and personal events.  Here, each person's fortune depends on the positions of the celestial bodies at their birth.  Babylonian associations between celestial and human events provided the foundation for astrology which, like the days of the week, endures today.  The cuneiform Venus Tablet of Ammisaduqa from c.1700 B.C. presents omens such as \cite{Aaboe01}:
\begin{quote}
[I]n month XI, the 18th day, Venus became visible in the east: springs will open, Adad will bring his rain, Ea his floods; king will send messages of reconciliation to king.
\end{quote}

Similar links between the deity, celestial and terrestrial affairs were present in ancient Judaism, a monotheistic culture with Mesopotamian heritage.  Psalm 8  from the Davidic era (c. 1000 BC) draws a link between God's power as creator of the celestial bodies and the human dominion over the terrestrial world \cite{Hobbins07}:
\begin{quote}
Yahweh our Lord, 
how magnificent is your name ... \\
When I witness your heavens
the work of your fingers \\
the moon and stars 
that you set in place \\
What are mortals
that you mind them ...\\
Yet you made them little less than divine \\
with glory and honor adorned them\\
you gave them reign over your hands' work.
\end{quote}

The intertwining of astronomy with religion and human affairs was sundered in the cosmology of Aristotle which dominated European thinking from Hellenistic through Renaissance times.  
Here the immutable heavenly realm was totally separated from the mundane sublunary realm.  The `quintessential' substance of the celestial realm had no relation to the substance of terrestrial world $-$ Earth, Air, Fire and Water.  In traditional Christian cosmology, which incorporated much of Aristotelian science, humanity reigned over the terrestrial world and, with sufficient faith and grace, could attain the heavenly realm after death.  Both astronomical study of the heavenly objects and astrological connections between their positions and terrestrial events were discouraged.  As a consequence, during the Middle Ages, astronomical research was far stronger under the Islamic empires than in Christian Europe.  

In this context, it is interesting to note that cosmologies integrating astronomy with terrestrial sciences did arise in the Hermetic tradition outside of church teachings.   Theophrastus of Hohenheim, or Paracelsus, in the 16th century was a leading figure who established important principles of modern chemistry and medicine \cite{Webster08}.  He developed a cosmology known as {\it Harmony of the Universe} where celestial bodies were associated with (al)chemical substances and the condition of human organs (Table 1).  Thus, specific diseases could be caused by planetary positions and might be cured by application of appropriate metals.  
\begin{table}[h]
\centering
\begin{tabular}{|lll|} \hline
{\bf Astrology} & {\bf Alchemy} & {\bf Medicine} \\ \hline
Sun & Gold & Heart \\
Moon & Silver & Brain \\
Jupiter & Tin & Liver \\
Venus & Copper & Kidneys \\
Saturn & Lead & Spleen \\
Mars & Iron & Gall bladder \\
Mercury & Quicksliver & Lungs \\ \hline
\end{tabular}
\caption{Scientific framework of Paracelsus}
\end{table}

Today, astronomy has strong connections to other sciences, as outlined in the following sections.  Unlike Aristotleian science, modern astronomy understands that the substance (e.g. atoms) and processes (e.g. gravitational force) in celestial objects is the same material as in the terrestrial world.  Except for the newly recognized Dark Matter and Dark Energy inferred from astronomical research but not yet discovered in laboratories, this represents an integration of astronomy with the terrestrial sciences of physics, geology, chemistry and mathematics.   

Unlike earlier times, a strong barrier now separates astronomy from commonly held religious and philosophical beliefs (except for the continuing belief in astrology which has little evidential foundation).  For example, galaxies or extrasolar planets are not associated with deities.  But there is a real potential for a re-integration of astronomy with broader culture.  Among all fields of physical science, astronomy is most widely known and enjoyed in media such as science fiction books, Hollywood movies, new magazines, and popular images.  

\section{Astronomy and statistics}

The field of statistics can be viewed as the mathematics of presenting and analyzing empirical data.  It also is a tool for scientific induction, allowing generalization about populations and processes in the physical world from limited observations \cite{Gregory05}:
\begin{quote}
The goal of science is to unlock nature's secrets. ... Our understanding comes through the development of theoretical models which are capable of explaining the existing observations as well as making testable predictions.  ... Fortunately, a variety of sophisticated mathematical and computational approaches have been developed to help us through this interface, these go under the general heading of statistical inference.
\end{quote}
The application of statistics to scientific data is not a straightforward mechanical enterprise, but requires careful statement of the problem, formulation of models, choice of statistical methods, calculation of statistical quantities, and judicious evaluation of the result.  Modern statistics is vast in its scope and methodology, so it is difficult to choose optimal paths for interpretation and inference.  

Astronomy, as the oldest observational science, played an unusually important role in the development of statistical methodology in past centuries \cite{Stigler86}.  From Greek times to the present, procedures for obtaining the `best'  estimate from discrepant astronomical observations have been debated.  Hipparchus advocated choosing the middle of the two extreme values while Ptolemy and al-Biruni recommend the mean of the extremes.  Some medieval scholars recommended that only one measurement be made, fearing that errors increased rather than decreased as repeated measurements were made.  Galileo provided a more modern discussion in his his 1632 book {\it Dialogue on the Two Great World Views, Ptolemaic and Copernican}.  He recommended using the mean of all observations, and he outlined in non-mathematical language many of the properties of observational errors later incorporated by Gauss into his quantitative theory.  

Newton's theory of gravity provided a new understanding of the motions of Solar System bodies which in turn raised new challenges in the application of mathematical formulae to discrepant observations \cite{Hald98}.  In the early 19th century, Laplace and Legendre developed methods for parameter estimation involving the minimization of the largest absolute residual or the sum of square residuals between the data and model.  The least squares method was placed on a probabilistic foundation by Gauss using the normal (or Gaussian) theory of errors. Least squares computations rapidly became the principal interpretive tool for scientific observations, and Gauss' distribution was widely known as the `astronomical error function'.  During the latter part of the 19th century, many leading astronomers contributed to least squares theory.  

However, from the late-19th through the late-20th centuries, the close collaboration between astronomy and statistics dramatically declined.  Statisticians concentrated on human sciences: demography, economics, medicine, psychology, and industries such as agriculture and manufacturing.  Astronomers, already dependent on Newtonian gravity, increasingly recognized the explanatory power of modern physics: electromagnetism, thermodynamics, quantum mechanics, and relativity.  

Thus, astronomers and statisticians substantially broke contact.  By the late 20th century, the curriculum for training astronomers was heavily dominated by physics with few if any courses in statistical theory or application.  Astronomers learned basic statistical methods, such as the least squares method, from practical guides \cite{Bevington69, Press86}.  Texts in statistics used examples from the biological or human sciences, rather than from the physical sciences.  

In recent decades, astronomers have increasing recognized the need for the sophisticated tools of modern statistics.  The challenges faced in modern astronomy span an amazing range of subfields. Whenever a table is presented with measured properties (in columns)  for a sample of objects (in rows), multivariate analysis is needed.  If the sample is heterogeneous, then multivariate classification is involved.  If variability in some property is studied, then time series analysis is required.  If the spatial distribution of objects is present (e.g., photons on a 2-dimensional image, galaxies in 3-dimensional redshift space, stars in 6-dimensional astrometric space), then density estimation and spatial point processes are involved.  If the data are fit to a parametric function, either a heuristic formula like a powerlaw or a complicated function derived from astrophysical theory, then regression and parameter estimation is pursued.  Several approaches to parametric modeling are used $-$ least squares, maximum likelihood, and Bayesian.  Methodological issues in model selection and validation are subtle and may be widely debated.  

An example of a topic in astrostatistics is the treatment of samples (e.g., stars, galaxies, sources at various wavelengths) which represent only tiny fractions of the underlying populations in the nearby universe.  Statistical inference is needed to make the leap from samples to populations.  Flux limits at the telescope lead to non-detections of known sources, known as left-censoring in statistical jargon, or entirely missing sources, known as truncation.  The field of `survival analysis', developed for biomedical research and industrial quality assurance, was found to be effective in treating censored data \cite{Feigelson85, Schmitt85}.  However, astronomical non-detections were tied to heteroscedastic measurement errors (i.e., different for each observation), and an integrated treatment of uncertainties and upper limits has only recently begun \cite{Kelly07}.  Statisticians had little experience with truncation, and major advances were made within the astronomical community.  Lynden-Bell (1971) derived the unique maximum-likelihood nonparametric estimator for randomly truncated data; its mathematical properties have been studied by statisticians \cite{Woodroofe85, Chen95}.  Methodological developments for estimating luminosity functions from truncated data is continuing today \cite{Schafer07, Kelly08}.   

The current state of statistical usage in astronomy is mixed.  One one hand, the average study uses only a narrow range of long-established methods: Fourier transforms (Fourier, 1809) for temporal analysis; least-squares and $\chi^2$ minimization for model fitting (Legendre 1805, Pearson 1901); Kolmogorov-Smirnov tests for goodness-of-fit measures (Kolmogorov 1933); and principal components analysis for multivariate structure (Hotelling 1936).  A myriad methods developed in recent decades are not widely known among astronomers. Even traditional procedures are often misused: incompatible least-squares fits are often viewed interchangeably \cite{Feigelson92}; the likelihood ratio test for comparing two models is often applied outside its allowed range \cite{Protassov02}; and the Kolmogorov-Smirnov goodness-of-fit test is used instead of the more sensitivie Anderson-Darling test \cite{Hou09}.  

On the other hand, the frontier of methodology in astronomical research has an active vanguard of experts, both in applying existing advanced statistics and in developing new methods.  A particularly strong interest in Bayesian modeling is rapidly emerging, particularly in cosmology \cite{Hobson09}.  A number of long-lived collaborations between astronomers and statisticians are now active such as the California/Harvard/ASC AstroStatistics Collaboration, International Computational Astrostatistics Group, and the Center for Astrostatistics. Conferences such as the {\it Statistical Challenges in Modern Astronomy} series bring experts from both fields together.  Summer schools for brief training in statistical principles and software have emerged on four continents.  The recent emergence of {\bf R} (http://r-project.org) as the dominant  public domain software system for statistical analysis and visualization is an important development, as astronomers had been hampered by a reluctance to purchase expensive proprietary systems designed for other communities.  

\section{Astronomy and chemistry}

Optical emission is characteristically produced by gases at temperatures $T > 3000$ K, so it was thought that the baryonic universe was mostly in the form of atomic and ionized gases. It slowly became evident through the 20th century that a cold component of molecular gases were also present in the Galactic interstellar medium.  Most of these molecules are identified by emission or absorption lines from quantized rotational transitions in the millimeter and submillimeter bands.  The hydroxyl molecule OH was first identified \cite{Weinreb63}, followed by ammonia NH$_3$, water H$_2$O, formaldehyde H$_2$CO and carbon monoxide CO.  Today, over 150 molecular species have been detected in interstellar clouds and circumstellar envelopes \cite{Woon09}.  Two Nobel laureates and authors of major monographs on molecular spectroscopy, Gerhard Herzberg and Charles Townes, were actively involved in the discovery of interstellar molecules.  As molecular species proliferated, understanding of astrochemical processes developed.  Processes such as charge-exchange and ion-molecular reactions, reactions catalyzed on solid grains at $T \sim 10$ K, non-equilibrium reactions in molecular clouds and protoplanetary disks are very unusual from the perspective of terrestrial chemistry.  

The discovery of cyclopropenyldiene, the first interstellar organic ring molecule, illustrates the intimate interactions between submillimeter astronomical observers, laboratory physical chemists, and theoretical quantum chemists \cite{Thaddeus85, Vrtilek87}.  A strong unidentified emission line had been seen at 85.338 GHz since 1976 in dense molecular cloud cores and circumstellar envelopes.  An absorption line at this frequency was finally detected in C$_2$H$_2$ laboratory discharge, but it was unclear what molecule had temporarily formed to produce the line.  A difficult trial-and-error process to find additional lines from the same molecule both in the ISM and laboratory succeeded with the discovery of a doublet at 184.328 GHz correlated with the 85 GHz line. This approximately defined the bond lengths and allowed calculation of other transitions of possible molecular carriers using a sophisticated quantum chemical model. The most promising molecule was the remarkable 3-carbon radical ring, c-C$_3$H$_2$ or cyclopropenylidene (Figure 1). Dozens of other emission and absorption lines from this molecule were quickly found in both the laboratory and at the telescope, confirming the identification.  

\begin{figure}[h]
\centering
\includegraphics[width=0.4\textwidth]{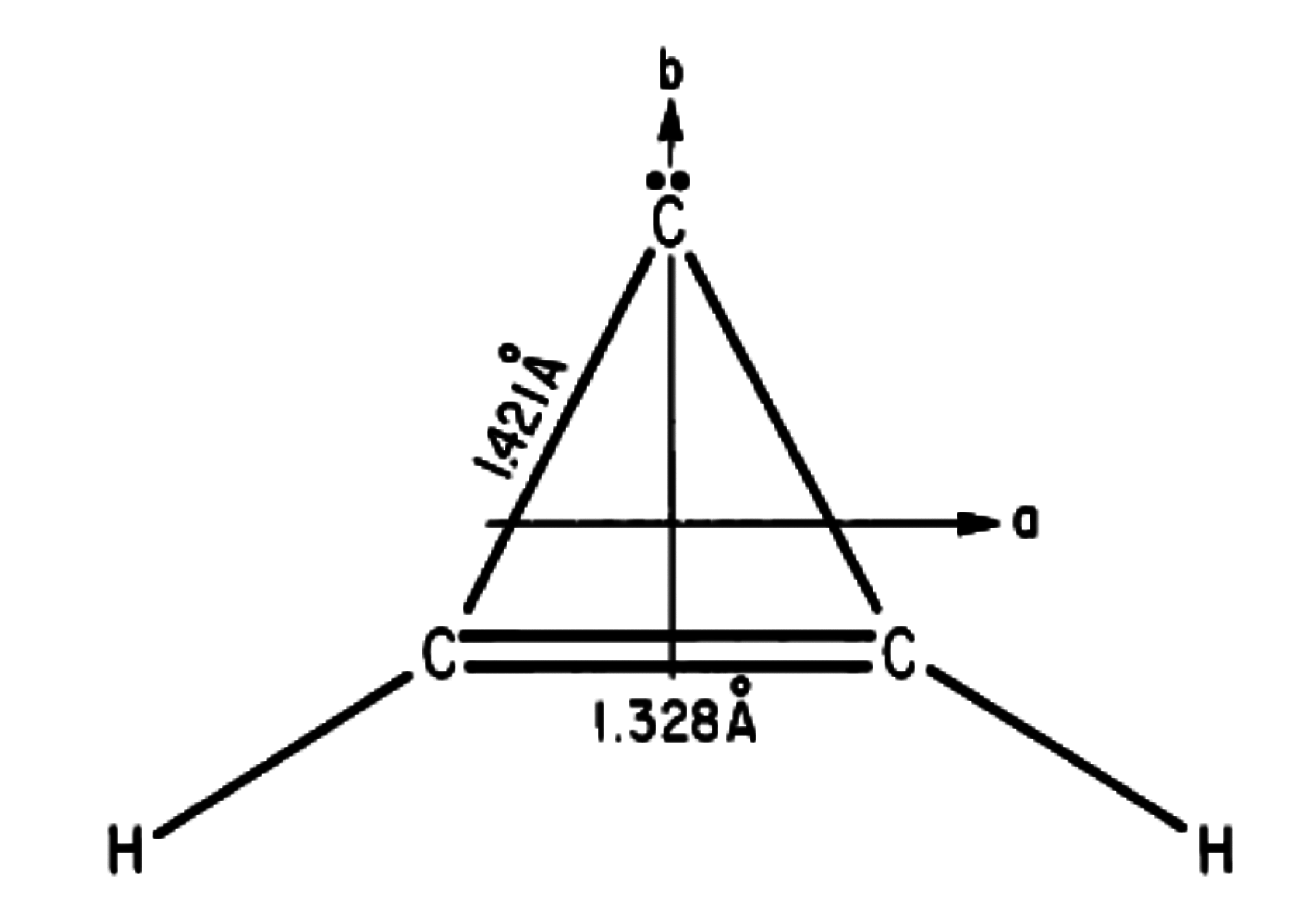}
\caption{Structure of the interstellar organic ring molecule cyclopropenylidene, c-C$_3$H$_2$.}
\end{figure}  

Another example of effective cross-disciplinary interaction in astrochemistry was the introduction of insights from hydrocarbon combustion into the explanation for the formation of interstellar polycyclic aromatic hydrocarbons or PAHs.  A significant fraction of the bolometric luminosity of dusty galaxies is emitted in several broad emission bands centered around 3.3,  6.2, 7.7, 8.6 and 11.3 $\mu$m.  Originally nicknamed the `unidentified infrared emission bands', their association with PAH molecules, essentially small 2-dimensional components of graphite with hydrogen atoms along the outer edges, was facilitated by comparing the astronomical spectra with the Raman spectrum of automobile exhaust \cite{Allamandola85}.  A chemist with expertise in soot formation in terrestrial hydrocarbon combustion asserted that PAHs should not readily form in astronomical environments because hydrogen atoms suppresses PAH synthesis at traditional soot forming temperatures, $T > 1400$ K.  However, he found that PAH formation was possible in H-rich atmospheres due to a balance of thermodynamic and kinetic effects which makes an acetylene addition step irreversible.  This occurs in a previously unrecognized narrow temperature window around $900 < T < 1100$ K, and is the likely mechanism for PAH formation in carbon-rich red giant stellar winds \cite{Frenklach89}.

\section{Other cross-disciplinary fields}

{\bf Physics}~~ As mentioned in \S 2, during the 20th century astronomy and physics became so closely wedded that astrophysics was often viewed as a branch of physics and astronomy was viewed as the empirical arm of astrophysics.  Consider the structure and evolution of stars, perhaps the greatest achievement of 20th century astrophysics.  Thermodynamics and gravity establish the overall equilibrium structure, nuclear physics determine the energy generation in the core, atomic physics regulates the transfer of energy through the interior and the spectral lines at the photosphere.  Magnetohydrodynamics and relativity have smaller effects in ordinary stars, but are critical for compact stars.   Study of the Sun had a major impact on particle physics with the solution to the `solar neutrino problem'.  

In other areas of astronomy, such as extragalactic astronomy, the observations give less detail and only relatively simple physical processes are sufficient to address the astronomical data.  Perhaps the most exciting interaction today between astronomy and physics is in cosmology.  High-precision measurement of the cosmic microwave background, galaxy clustering structure, and hot plasma outside of galaxies are in remarkable agreement with a an astrophysical model of cosmic expansion dominated by Dark Matter and Dark Energy.  

{\bf Engineering}~~ Perhaps second only to physics, and less widely discussed, is the powerful dependence of astronomical discoveries on progress in applied physics and engineering.   CCD detectors, infrared detector, adaptive optics and satellite technologies all have a strong foundation in military technologies.  High electron mobility transistor technology, developed at Fujitsu and Bell Labs in the 1970s are the critical technology for receivers on the forthcoming Atacama Large Millimeter Array.  Corning and Hughes Danbury optics factories in the U.S. produced mirrors for the Hubble Space Telescope, Chandra X-ray Observatories, Gemini and Subaru telescope.  Modern telescope systems can not have been built without access to rapidly improving technological capabilities.

{\bf Earth sciences}  ~~ The study of Solar System planets and minor bodies is conducted more in the context of Earth sciences $-$ geology, atmospheric studies and related fields $-$ than of astronomy.  But with the recent discoveries and characterizations of exoplanets, collaborations are growing between astronomers and Earth scientists in the understanding of planetary atmospheres, surfaces and chemistry.   Several new cross-disciplinary collaborations, similar to those in astrostatistics, have emerged to address these issues.

The study of meteorites, rocks from the ancient Solar System impacting Earth, has perhaps been the most fruitful area of interplay between geology and astronomy.  Remarkably detailed laboratory study of isotopic abundances in selected grains of pristine meteorites reveal patterns that can be convincingly attributed to interstellar grains formed in the atmospheres of red giant stars \cite{Lewis90, Gallino90}. Meteoritics and cometary studies in our Solar System are now richly interacting with the mineralogy of protoplanetary disks revealed by infrared spectroscopy to unravel the complex processes involved in the origin of planets \cite{Reipurth07}.  

{\bf Informatics}  ~~ A nascent interest in advanced computational methods and informatics is emerging among some astronomers.  A few projects are well-established.  The SIMBAD and Vizier services of the Centre de Donn\'ees Astronomique de Strasbourg (CDS) provides access to a wide range of tabular and bibliographic data in useful formats.  These capabilities required synergy between astronomers, specialized librarians and computer engineers.  The NASA/Smithsonian Astrophysics Data System, a billion-hit Web site, has become the principal portal into the astronomical literature.  It has become increasingly intertwined with electronic services of the main journals, the CDS, and some observatory archives.  

Astroinformatics is being propelled by a variety of projects generating exceedingly large and complex datasets.  The International Virtual Observatory (IVO) is an ambitious effort to federate widely distributed online databases from many telescopes.  Assisted by computer scientists, the basic access tools are now in place, but the scientific usage of the IVO is just beginning.  The Large Synoptic Survey Telescope (LSST) is expected to generate many petabytes of data, essentially a video of the entire sky, and ancillary products such as multi-billion object catalogs.  New astronomical specialities developing high-performance computing (including distributed `cloud' computing), data mining, knowledge storage and representation, and semantic astronomy are emerging \cite{Borne09}.  Altogether, the new cross-disciplinary field of astroinformatics,  analogous to the more advanced fields of bioinformatics and geoinformatics, will grow rapidly in coming years.

{\bf Life sciences}  ~~ Astrobiology, the study of life in the universe, is one of the most intriguing new cross-disciplinary fields of astronomy.   In some sense, it does not yet have an empirical basis as there is no evidence for lifeforms outside of Earth.  But many of the elements underpinning such a discover are rapidly progressing.  Biological and genetic study of extremophiles and Archaea reveal the versatility of lifeforms on early Earth.  {\it In situ} and remote studies of Mars, Jovian moons, asteroids and comets give insights into chemical, mineralogical, atmospheric and climatic conditions relevant for the emergence of life.  Discovery of extrasolar planets elucidate the complexities of planetary system formation processes.  A number of planets are already known in the Habitable Zones around nearby stars, and major specialized space-borne telescope systems will enormously increase current samples. Studies of solar and stellar activity give external conditions which affect planetary atmospheres necessary for life. Radio surveys of nearby stars in SETI programs provide a chance for revolutionary discovery of technologically advanced life.   

\section{Do cross-disciplinary approaches advance astronomical discovery?}

From the brief review above, it is clear that cross-disciplinary approaches have significantly contributed to the advancement of astronomical insights in the past, and the prospects for future roles are bright.  Cross-disciplinary researchers consistently make profound, revolutionary transformations in our thinking.  Astronomy would obviously be impoverished without its symbiosis with physics since Newton's foundation of celestial mechanics and the application of quantum mechanics to stellar spectra \cite{Payne25}.  But consider how our understanding today of molecular processes in cosmic environments would be impoverished had leaders like Herzberg and Townes not brought their expertise to the astronomical community fifty years ago.    Collectively, the roles of astrochemistry, astrostatistics, astrogeology, astroinformatics,  astrobiology and astronomical instrumentation have been enormous in furthering astronomical understanding.  

These cross-disciplinary fields contribute to astronomy in two ways:  directly making discoveries outside the core field of astronomy, and enhancing discoveries within the core.  In astrochemistry, research in physical and quantum chemistry was essential for the 
discovery of complex organic interstellar molecules.  Close cross-disciplinary collaboration between the microwave-band astronomer, laboratory chemist and theoretical quantum chemist (who may or may not be separate individuals) is needed. In engineering for instrumentation, astronomical discovery in entire fields such as far-infrared and gamma-ray astronomy depend on high-technology detectors and on space-flight technologies to lift the telescope above the Earth's obscuring atmosphere.  The replacement of photographic plates with CCDs in the 1980s and the improvements of correlators in radio interferometers in the 2010s are examples where engineering advances have increased sensitivities of established telescopes by an order of magnitude.  

On the other hand, astrostatistics and astroinformatics do not direct discovery but rather improve the quality of our understanding of complex datasets.  The philosopher Ian Hacking has written: "The quiet statisticians have changed the world, not by discovering new facts ...  but by changing the ways that we reason, experiment, and form our opinions".  Relatively unimportant in traditional astronomy, these skills become essential for tapping the insights in megadatasets from large surveys and in precision model fitting.  For example,  advanced Bayesian statistical techniques are used to derive cosmological parameters in $\Lambda$CDM astrophysical models using large high-precision datasets from the WMAP and Planck satellite-borne microwave telescopes \cite{Hobson09}.

\section{The future of cross-disciplinary astronomy}

Despite the successes of cross-disciplinary research in astronomy, there are structural difficulties that can be $-$ and should be $-$ ameliorated to maximum future scientific return.  It depends mostly on broadening the attitude of majority of core astronomers to appreciate and support the minority of cross-disciplinary astronomers.  Two specific avenues for improvement are needed.  

First, young astronomers must be taught more broadly and flexibly, with meaningful exposure to engineering, statistics, chemistry, geology (and perhaps biology) as they apply to astronomical problems.  At present, the curriculum for training research astronomers concentrates on a strong background in physics (and associated background in mathematics) with little or no opportunity for the interested student to pursue cross-disciplinary interests.  Informal training in summer schools, and speciality texts for self-instructions are also valuable elements for education in these areas.  

Second, sufficient and stable research funding is essential for cross-disciplinary research.  The criteria for success in these fields may differ from those in core astronomy; the major products may be a software system or a physical chemistry laboratory rather than rapid publications in the primary astronomical journals.  Effective evaluation of research proposals is difficult;  there is a tendency for disciplinary panels to downweight unfamiliar cross-disciplinary efforts.  Observatory leaders may also be reluctant to devote scarce resources to expensive personnel with credentials outside the core areas.  The funding of astrostatistics, for example,  is far below the needs of the field, and far below the funding devoted to methodology in the Earth and biological sciences.  There is also a worry that physical chemists are receiving insufficient funding to identify molecular carriers of the many lines from the revolutionary submillimeter spectrometers on the Herschel and ALMA telescopes.  

If the paths of cross-disciplinary astronomy are groomed and talented scientists tread them in a spirit of creativity, then we can envision a blossoming of astronomy in new areas.   The possibility of astronomy leading a broader alliance with non-scientific culture, similar to the traditional integrated worldviews of ancient societies, should not be ignored.  The emergence of syncretic churches with Ph.D. priests promulgating religious texts based on modern astrophysics is not unimaginable.  The recent book {\it View from the Center of the Universe} by Primack and Abrams \cite{Primack06}, a research cosmologist and a cultural philosopher respectively, provides a fascinating linkage between modern astronomy and religious symbols.  

{\it Acknowledgements:}  I would like to thank Michael Frenklach for 5 years of stimulating collaboration in astrochemistry, G. Jogesh Babu for 20 years of exciting collaboration in astrostatistics, and my wife Zo\"e Boniface for 25 years of enthusiasm for my cross-disciplinary activities.

\end{document}